\documentclass[preprint,amsmath,amssymb,prd]{revtex4}
\usepackage{epsf}
\usepackage{amsmath,amsthm,amssymb}
\usepackage{color}
\usepackage{dcolumn}
\usepackage{bm}
\usepackage{graphicx,epsfig}
\usepackage[dvipsnames]{xcolor}
\usepackage{ulem}

\graphicspath{{fig/}}
\everymath{\displaystyle}

\begin{document} 

\title{Production of twisted particles in heavy-ion collisions}

\author{Liping Zou$^{1}$}
\email{zoulp5@mail.sysu.edu.cn}
\author{Pengming Zhang$^{2}$}
\email{zhangpm5@mail.sysu.edu.cn}
\author{Alexander J. Silenko$^{3,4,5}$}
\email{alsilenko@mail.ru}

\affiliation{$^1$Sino-French Institute of Nuclear Engineering and Technology, Sun Yat-sen University,
519082 Zhuhai, China}
\affiliation{$^2$School of Physics and Astronomy, Sun Yat-sen University,
519082 Zhuhai, China}
\affiliation{$^3$Bogoliubov Laboratory of Theoretical Physics, Joint
Institute for Nuclear Research, Dubna 141980, Russia}
\affiliation{$^4$Institute of Modern Physics, Chinese Academy of
Sciences, Lanzhou 730000, China}
\affiliation{$^5$Research Institute for Nuclear Problems, Belarusian
State University, Minsk 220030, Belarus}

\begin{abstract}
A prevalence of production of twisted (vortex) particles in noncentral heavy-ion collisions is shown. In such collisions, photons emitted due to the rotation of charges are highly twisted. Charged particles are produced in nonspreading multiwave states and have significant orbital angular momenta. It can be expected that an emission of any twisted particles manifesting themselves in specific effects is rather ubiquitous.
\end{abstract}

\maketitle
\section{Introduction}
A twisted (vortex) particle possesses an intrinsic orbital angular momentum (OAM) and can be presented by a wave beam or a packet. Wave beams and packets are localized with respect to two and three dimensions, respectively and are characterized by two discrete transverse quantum numbers. Free twisted particle beams of photons, electrons, and neutrons can be described in the cylindrical coordinates by the Laguerre-Gauss (LG) wave function
\cite{Kogelnik,Siegman,Alda,Pampaloni} .
Free twisted particles can possess huge intrinsic OAMs. Photons with OAMs more than 10000$\hbar$ \cite{PNAS} and electrons with OAMs up to 1000$\hbar$ \cite{VGRILLO} have been obtained.
Twisted photons have unusual properties like a subluminality and nonzero effective masses \cite{photonPRA}.  The general quantum-mechanical solutions for twisted electrons in a uniform magnetic field, as has been recently shown in \cite{arXiv}, have the form of LG beams undergoing spatial oscillations and containing the Landau states as a specific case.
 Charged twisted particles can be recognized by their dynamics \cite{Manipulating,ResonanceTwistedElectrons}, magnetic moments \cite{BliokhSOI}, and specific effects in external fields \cite{KarlovetsZhevlakov,PhysRevLettLanzhou2019,snakelike,arXiv} and collisions \cite{OAMinteraction}.

The importance of production of twisted particles in heavy-ion collisions (HIC) is evident. The similar effect of the global polarization of produced particles \cite{GlPolar} attracts a lot of attention. To study the above-mentioned problem at noncentral HIC, one should take into account an appearance of a strong magnetic field which is generally a consequence of strong interaction \cite{SkokovIllarionovToneev}, and a fast rotation of nuclear matter leading to its large vorticity \cite{Baznatetal}.
In this paper, we consider the possibilities to find twisted particles in the highly nontrivial environment provided by HIC, and hereafter the strong magnetic field and large vorticity will be regarded as two independent inherent properties of HIC. 
\\
\section{Twisted photon in HIC}
The recent theoretical \cite{Katoh,Epp,theoretical} and experimental \cite{experimental} papers unambiguously show that the photons radiated by electrons in a helical or circular motion 
are twisted (i.e., have nonzero OAMs). Similar effects takes place for rotating black holes \cite{BlackHoles,BlackHoleE}. Since electrons move in the magnetic field $B$, parameters of synchrotron radiation are expressed in terms of $B$. However, the synchrotron radiation is caused by an electron acceleration and is defined by an electron motion. Its dependence on $B$ is conditioned by only the fact that the electron motion is defined by the magnetic field. Certainly, the synchrotron radiation can de described in terms of the velocity and acceleration of rotating charges. Therefore, a rotation of charges at HIC should lead to a photon production equivalent to the synchrotron radiation. This radiation is defined by the rotation of charges and is independent of the magnetic field which also appears due to the above-mentioned rotation.

The synchrotron radiation spectrum consists of frequencies multiple to the rotation frequency of the compound nuclei, $\Omega$:
\begin{equation}
\omega_j=j\Omega,\qquad j=1,2,3,\dots
\label{mflB}
\end{equation} Amazingly, the total angular momentum of a radiated photon is equal to $\hbar j$ and has orbital and spin parts \cite{Katoh,Epp}. All photons with $j>1$ are twisted. This effect confirmed by experimental data \cite{experimental} clearly shows that twisted photons are widespread in nature.

Massive particles can also be produced in twisted states. The vorticity of nuclear matter is, to some extent, an effect which is similar to the twist and leads to a production of particles in twisted states \cite{footnote}. To describe processes taking place in a rotating nuclear matter, one often uses uniformly rotating frames (see, e.g., Ref. \cite{Fukushima:2020ncb,Huang:2018aly,Liu:2017spl,Jiang:2016wvv,Chen:2015hfc}). For a quantum-mechanical description of particles in such frames, it is convenient to apply the Foldy-Wouthuysen (FW) representation providing for the Schr\"{o}dinger picture of relativistic quantum mechanics (see Refs. \cite{PRA2015,PRAFW} and references therein). The relativistic FW Hamiltonian describing a spin-1/2 particle in the frame uniformly rotating with the angular velocity $\bm \omega$ has the form \cite{PRD2}
\begin{equation}
\begin{array}{c}
{\cal H}_{FW}=\beta\sqrt{m^2+\bm p^2}-\bm \omega\cdot(\bm L+\bm s),
\end{array}\label{eqrotfr}
\end{equation}
where $\beta$ is the Dirac matrix and $\bm L$ and $\bm s$ are the OAM and spin operators. The eigenstates of this operator have definite integer OAMs and the corresponding eigenfunctions describe LG beams.

In HIC of Au-Au 
(RHIC) and Pb-Pb 
(SPS and LHC) nuclei,
the collision energies $\sqrt {s_{NN}}$ are up to 200 GeV, 17.2 GeV, and 5.02 TeV, respectively \cite{AuAu1,AuAu2,AuAu3,RHIC,PbPb-sps1,PbPb-sps2,PbPb-LHC1,PbPb-LHC2,PbPb-LHC3,Noferini}. In the collision picture, nucleus are nonspherical and are mostly like a disc \cite{Schuetrumpf}. The $_{79}$Au$^{197}$ nucleus radius is equal to $R=6.98$ fm and the Coulomb energy of two touching Au nuclei is 
$W=Z^2e^2/R=1.29$ GeV.
Evidently, the Coulomb repulsion of colliding nuclei can be neglected.

Let us consider the collision with the impact factor $b=4$ fm. We can determine the momentum components which are normal and tangent to the plane tangent to the surfaces of the two touching nuclei. We suppose that the normal and tangent momentum components of each colliding nucleus in the center-of-mass frame contribute to internal and rotational energies of the compound nucleus, respectively. Since these momentum components are defined by $p_n=p\sqrt{1-b^2/(4R^2)}\approx 0.71p,\, p_\tau=pb/(2R)\approx 0.29p$, the average velocities of rotating charges (protons or quarks) should be ultrarelativistic. It can be checked that a compound nucleus cannot rotate at HIC as a solid ($\Omega=const$). Since $v=\Omega r$, only thin external layers of solids ($R-r\ll R$) can be ultrarelativistic ($v\approx c$). Velocities of other layers are defined by $v\approx cr/R$. Such layers cannot be ultrarelativistic. For nucleons forming such layers, the average energy of a single nucleon  cannot be much bigger than $m_pc^2$. This result contradicts to the energy and angular momentum conservation because rotational energies of the colliding nuclei are ultrarelativistic. Therefore, all or almost all rotating charges should be ultrarelativistic. In this case, transverse momenta of usual partons within nucleons, $k_T$, should be rather significant. In the RHIC and LHC experiments, they are reported to be a few GeV \cite{ZEUS:2001lmk,PHENIX:2006gto,Mondal:2009mer,BermudezMartinez:2020tys,Bacchetta:2022awv}. However, it is difficult to establish a relationship between the quantities $p_\tau$ and $k_T$. We admit that a strongly interacting and rotating nuclear fluid can contain some vortices, especially in the case of an appearance of quark-gluon plasma. In this case, the radius of trajectory of rotating charges, $\rho$, can be significantly less than the nucleus radius. However, we suppose that $\rho$ cannot be less than the proton charge radius ($\rho\geqslant r_p$, $r_p=0.84$ fm \cite{ProtonRadius}) because confinement forces are unsufficiently strong on a smaller distance. 
For the quark-gluon plasma, some estimates can in principle be substantially different.

The synchrotron radiation spectrum is formed by lines which frequencies are harmonics of the rotation frequency [see Eq. (\ref{mflB})] and is
defined by well-known formulas \cite{LL2}. When the charges are ultrarelativistic, the spectral maximum in classical electrodynamics is defined by 
\begin{equation}
\omega\sim\Omega\gamma^3.
\label{clasleq}
\end{equation}
This relation determines a large orbital polarization of radiated photons.  However, quantum electrodynamics (see Eqs. (90.3) and (90.24) in Ref. \cite{LL4}) leads to the different connection between the spectral maximum $\omega$ and the particle energy 
$\epsilon=\gamma m$:
\begin{equation}
\begin{array}{c}
\frac{\hbar\omega}{(\epsilon-\hbar\omega)\chi}\sim1,\qquad \chi\approx
\frac{\hbar\Omega{\epsilon}^2}{m^3}.
\end{array}\label{QED}
\end{equation}
Our denotations almost coincide. $\omega_0$ in Ref. \cite{LL4} is our rotation frequency $\Omega$, and the substitution $\epsilon'=\epsilon-\hbar\omega$ is made. The classical theory can be used, when the photon energy $\hbar\omega$ is small as compared
with the particle energy. When $\hbar\omega\approx\epsilon$ and $\epsilon-\hbar\omega\ll\epsilon$, we obtain the evaluation of the photon OAM  $L\approx\hbar\omega/\Omega\approx\gamma m/\Omega$. To evaluate this quantity in the general case, one can apply the classical equation (\ref{clasleq}) but should take into account the quantum-mechanical cutoff:
\begin{equation}
\begin{array}{c}
L\sim\min{\left(\hbar\gamma^3,\frac{\gamma m}{\Omega}\right)}.
\end{array}\label{general}
\end{equation}
For ultrarelativistic protons and valence quarks, $\Omega\sim c/\rho$. The ratio 
$m/(\hbar\Omega)$ is maximum in the case of $m=m_p,\,\rho=R$ and is approximately equal to 33.2. For confined constituent quarks, $m=m_q=m_p/3,\,\rho=r_p$, and the ratio $m/(\hbar\Omega)$ is approximately equal to 1.3. As a result, $L$ is large enough ($\gamma\gg1$) and
photons emitted due to electromagnetic interactions at noncentral HIC are significantly twisted. Thus, our model-independent analysis rigorously proves a significant orbital polarization of such photons.

\section{Angular momentum of charged twisted particles in HIC}\label{OAMinHIC}
The following analysis based on precedent studies \cite{Gori,Brinkmann,Edwards,Burov,Kim,Sun,Crewe,Wakamatsu,FloettmannKarlovets,Karlovets2021} allows us to predict that charged particles produced in 
noncentral HIC also have significant OAMs. It is very important to determine what quantum states are suitable for a production of such particles. Hereinafter, basic 
and nonbasic states denote quantum states suitable and unsuitable for such a production \cite{ffootnote} and  $\bm\pi=\bm p-e\bm A$ is the kinetic momentum.
Basic states in the quantum-mechanical description of charged particles in a uniform magnetic field are the Landau states with a definite longitudinal momentum. In the Landau states, the canonical (generalized) 
and kinetic (mechanical) OAMs ($\bm L=\bm r\times\bm p$ and $\boldsymbol{\mathcal{L}}=\bm r\times\bm\pi$, respectively) can be not only parallel but also antiparallel.
The canonical OAM is conserved. In a uniform magnetic field with the symmetric gauge of the vector potential $\bm A=\bm B\times\bm r/2$, its connection with the kinetic OAM 
$\boldsymbol{\mathcal{L}}=\bm r\times\bm\pi$ is given by (see Refs. \cite{Wakamatsu,FloettmannKarlovets})
\begin{equation}\begin{array}{c}
\boldsymbol{\mathcal{L}}=\bm L-\frac e2\bm r\times(\bm B\times\bm r)=\bm L-\frac {e\bm B}{2}r^2,
\end{array}\label{numberd}
\end{equation}
where $r=|\bm r|$ is the radial coordinate in the plane orthogonal to $\bm B$. The kinetic OAM, unlike the canonical one, is gauge-invariant (see Ref. \cite{Wakamatsu}). We denote $\bm B=B\bm e_z$ and $L_z\equiv\ell\hbar$. Relativistic wave functions in the FW representation defining the Landau states are based on the nonrelativistic 
solution \cite{Landau,LL3} and are presented, e.g., in Refs. 
\cite{JINRLETTERS2008,paraxialLandau}. 
For relativistic particles with a negative charge, the Landau energy levels are defined by (see Ref. \cite{paraxialLandau} and references therein)
\begin{equation}
\begin{array}{c}
E=\sqrt{m^2+p_z^2+(2n+1+|\ell|+\ell+2s_z)|e|B},
\end{array}
\label{eqOAM}
\end{equation} where $n=0,1,2,\dots$ is the radial quantum number. For relativistic particles with a positive charge, $\ell$ in this equation should be replaced with $-\ell$.
The \emph{exact} relativistic solution of the problem 
shows the infinite degeneracy of the Landau states with ${\rm sgn}(e)\ell>0$ and an arbitrary $n$. 

It has been shown in Ref. \cite{JINRLett12} with the use of the Wentzel-Kramers-Brillouin method that the classical limit of \emph{relativistic} quantum-mechanical equations is reduced to the replacement of operators in the Hamiltonian and quantum-mechanical equations of motion by the respective classical quantities.
In the classical picture, particles move in a helix and the radial motion does not occur. Well-known relativistic classical formulas defining the particle dynamics lead to the following relations:
\begin{equation}
r=|r|=-\frac{\pi_\phi}{eB},\qquad \mathcal L_z=2L_z=-e Br^2=-\frac{\pi^2_\phi}{eB}.
\label{OAMcl}
\end{equation}

The Landau state radii with respect to the cychrotron motion center are defined by \cite{integral,Li}
\begin{equation}
\langle r^2\rangle=\frac{2(2n+|\ell|+1)}{|e|B},
\label{radii}
\end{equation} Therefore, the connection between the kinetic and canonical OAM operators in the FW representation is given by (see Eq. (\ref{numberd}) where $\bm L$ is an invariant)
\begin{equation}\begin{array}{c}
\langle\mathcal{L}_z\rangle=\ell\hbar-{\rm sgn}(e)(2n+|\ell|+1)\hbar\\=L_z-{\rm sgn}(e)[(2n+1)\hbar+|L_z|],\\ \langle\mathcal{L}_z\rangle-2L_z=-{\rm sgn}(e)[2n+|\ell|+{\rm sgn}(e)\ell+1]\hbar.
\end{array}\label{ckOAM}
\end{equation} 
Due to the absence of the radial motion in the classical picture, Eqs. (\ref{OAMcl}) and (\ref{ckOAM}) agree for ${\rm sgn}(e)\ell\le0$ and disagree for ${\rm sgn}(e)\ell>0$. Thus, the Landau states with ${\rm sgn}(e)\ell\le0$ and ${\rm sgn}(e)\ell>0$ are basic and nonbasic for the particle production (see Ref. \cite{preprint} for more details). This conclusion does not mean that nonbasic states are unphysical. Particles can be created in the basic Landau states in the magnetic field $\widetilde{\bm B}\neq\bm B$ and then penetrate to the area of the field $\bm B$. 
After the penetration, their quantum states are described by the LG beams or, in the specific case of $\widetilde{\bm B}=-\bm B$, by the nonbasic Landau solution with ${\rm sgn}(e)\ell>0$. This result explains 
the division of the Landau states into basic and nonbasic.
Nonuniform magnetic fields cover uniform ones. Therefore, charged particles cannot be created in \emph{nonuniform} magnetic fields in quantum states with opposite directions of canonical 
and kinetic OAMs .

It has been demonstrated in Refs. \cite{FloettmannKarlovets,Karlovets2021} on the basis of the quantum Busch theorem \cite{Busch,Reiser} that particle beams generated inside a magnetic field take the form of LG beams 
outside of this field. This important achievement predicts wide dissemination of twisted beams in nature.
It has been noted in Refs. \cite{FloettmannKarlovets,Karlovets2021} that twisted particles are produced in states with $\mathcal{L}_z=0$ and $L_z\neq0$. This note is absolutely correct for specific experimental conditions 
\cite{FloettmannKarlovets,Karlovets2021} and does not contradict to our analysis.
For a real solenoid, it is convenient to choose the symmetric gauge with the solenoid symmetry axis corresponding to $r=0$.
When a particle is produced near this 
axis, $\mathcal{L}_z=0$ at the moment of creation and then 
increases or decreases up to the extreme value $-e Br^2=-\pi^2_\phi/(eB)$. Unlike Refs. \cite{FloettmannKarlovets,Karlovets2021}, we condider the case of coinciding the trajectory symmetry axis with the 
solenoid one.

However, for the geometry used in Refs. \cite{FloettmannKarlovets,Karlovets2021}, the vector potential gauge should be changed.
It can be shown that the average values $\langle\mathcal{L}_z\rangle$ depend on a shift of the former axis from the latter one.

We define the shift $\bm R_0$, and consider the radius of a circular motion $r$ of the particle which is much smaller than the solenoud radius. So we can use approximation $\bm B(\bm R)\approx\bm B(\bm R_0)$.  In this case, the vector potential is expressed in symmetric gauge as follows:
\begin{eqnarray}
\bm A(\bm R)=\bm A_0(\bm R_0)+\bm{\mathcal{A}}, \quad \bm A_0=\frac12\bm B(\bm R_0)\times\bm R_0, \quad
\bm{\mathcal{A}}=\frac12\bm B(\bm R_0)\times\bm r,\quad \bm R=\bm R_0+\bm r.
\label{eqelfll}\end{eqnarray}  
We consider particle position far from the solenoid edge, $B_z\approx B$, $A_{0\phi}\approx R_0B/2$, and $\mathcal{A}_\phi\approx rB/2$, and can check that  $\langle\bm r\rangle=0,\,\langle\bm{\mathcal{A}}\rangle=0,\,\langle\bm \pi_\bot\rangle=0$.

As a result, for electron production, we obtain
\begin{equation}
\mathcal{L}_z=L_z+\frac {|e|B}{2}\left[R^2_0+\frac{\pi^2_\bot}{e^2B^2}+2R_0\frac{\pi_\bot}{|e|B}\cos{(\omega t+\phi)}\right].
\label{equel}
\end{equation}
where $\omega$ is the particle rotation frequency (cyclotron frequency), and we can see here clearly the dependence of $\langle\mathcal{L}_z\rangle$ on the axis shift $R_0$.

Twisted electrons were produced by the photoemission in experiments at FNAL \cite{Sun} and had rather large OAMs of the order of $10^8\hbar$ in magnetic fields $B\sim10^{-2}-10^{-1}$ T.

It has been obtained in Ref. \cite{SkokovIllarionovToneev} that the magnetic field emerging in the SPS, RHIC, and LHC experiments is of the order of 0.1 $m_{\pi}^2/|e|$, $m_{\pi}^2/|e|$, and 15 $m_{\pi}^2/|e|$, respectively \cite{footnote2}.
The maximum beam radius is restricted by the area of the nonuniform magnetic field and is of the order of $r_{max}\sim 3\div10$ fm (see Ref. \cite{SkokovIllarionovToneev}). As follows from Eq. (\ref{radii}) at $n=n_{min}=0$,
\begin{equation}
L_z^{max}=\frac{|e|Br^2_{max}}{2}-\hbar.
\label{OAMmaxi}
\end{equation} For the above-mentioned experiments, the maximum values of $|\pi_\phi|$ and $|\ell|$ defined by Eqs. (\ref{OAMcl}), (\ref{OAMmaxi}) read 
$|\pi_\phi|_{max}\sim0.1\,{\rm GeV}/c,\,|\ell|_{max}\sim1$; $|\pi_\phi|_{max}\sim1\,{\rm GeV}/c,\,|\ell|_{max}\sim10$; and $|\pi_\phi|_{max}\sim10\,{\rm GeV}/c,\,|\ell|_{max}\sim10^2$, respectively. The total 
longitudinal momentum can be much bigger than $|\pi_\phi|_{max}$ if a produced particle is paraxial ($|\bm\pi_\bot|\ll|\bm\pi|$) relative to the magnetic field direction. Evidently, the orbital polarization of charged particles produced in HIC is
not negligible. Since it manifests in processes studied in Ref. \cite{OAMinteraction}, it can in principle be discovered. The vorticity of nuclear matter is another reason of a production of not only charged 
but also uncharged twisted particles \cite{footnote}.

The magnetic field at HIC is substantially nonuniform and time-dependent. The strong magnetic field \emph{near colliding ions} \cite{SkokovIllarionovToneev} is able to hold charged particles even with large transversal momenta (see the precedent paragraph) in a helix. 
However, a quick decrease of this field with an increase of a distance from colliding nuclei strongly hampers a motion of such particles in the helix and a detection of their OAMs. Therefore, we need to focus our attention on producing \emph{paraxial} charged particles $(|\bm\pi_\bot|\ll|\bm\pi|)$ having less transversal kinetic momenta. Even $(|\bm\pi_\bot|/|\bm\pi|)\lesssim0.1$ can often be sufficient.

\section{Discussion}
It has been also shown in Refs.
\cite{arXiv,FloettmannKarlovets,Karlovets2021} that twisted beams are not spreading in nonuniform electric and magnetic fields and their propagation does not destroy the beam coherence. Nevertheless, the beam coherence at HIC needs a separate analysis because relativistic effects and spin interactions should be considered. A rigorous relativistic FW transformation providing for the Schr\"{o}dinger-like form of quantum mechanics for particles with different spins and arbitrary energies in arbitrarily strong external fields can be fulfilled by the method developed in Refs. \cite{JMP,PRA2015,PRA2008}. This method allows one to calculate \emph{exactly} leading terms in FW Hamiltonians proportional to the zero and first powers of $\hbar$ and terms describing contact interactions which are proportional to $\hbar^2$ \cite{PRA2015}. The corresponding relativistic FW Hamiltonian has been obtained in Refs. \cite{PRAFW,JMP,RPJ}. It contains spin-independent and spin-dependent terms and is the same for untwisted and twisted spin-1/2 particles.

In general, particle wave beams penetrating to free space do not satisfy the paraxial approximation. In strong \emph{stationary} electric and magnetic fields, particles are produced in some quantum states with the fixed energy, ${\cal H}_{FW}\Psi_{FW}=E\Psi_{FW}$. These states generalize the Landau ones and the corresponding wave functions describe nonparaxial and paraxial twisted beams. The beams are formed by partial waves with the same constant energy $E$ which become plane waves in free space. Since the wave coherence in a given spatial point is provided for coinciding the wave energies, the energy conservation in stationary fields leads to the coherence of the partial waves in the whole space. Momenta of the partial waves do not coincide, can differently vary from point to point, and depend on the spin polarization. However, these properties do not violate the beam coherence. In principle, the beam shape near colliding ions can be deduced from its shape in the free space and slightly depends on the spin polarization.

Colliding ions are nonstationary quantum systems and their fields are not fully stationary. Nevertheless, this property does not violate the beam coherence in a given spatial point if the nonstationarity does not eliminate the equality of energies of the partial waves in this point. 

This problem should be considered in more details. To simplify the consideration, we disregard spin interactions. The beam rotation caused by the Lorentz force
\begin{equation}
\bm F_L=\frac{d\bm\pi}{dt}=e(\bm E+\bm v\times\bm B) \label{Lornf}
\end{equation}
vanishes the \emph{average} beam momentum in the plane orthogonal to the magnetic field. However, partial particle waves have nonzero transversal momenta. The Hamiltonian has the form
\begin{equation}
{\cal H}=\sqrt{m^2+\bm\pi^2}+e\Phi, \label{HamltFW}
\end{equation} where $\Phi$ is the scalar potential. The corresponding quantum-mechanical equations are the same but operator products should be presented in the Hermitian form \cite{JMP}. Since the force in any stationary and nonstationary magnetic field is orthogonal to the kinetic momentum, any magnetic field does not change the total beam energy. This energy is not changed by any stationary electric field and the longitudinal component of a \emph{nonstationary} electric field conditions equal changes of the total beam energy for all partial waves. Since transversal momentum components of partial waves have different directions, the transversal component of a nonstationary electric field differently changes transversal momenta and energies of partial waves. As a result, the difference of total energies and the noncoherence of partial beams appear. Nevertheless, this noncoherence seems to be rather small. The nonstationary electric field changes the transversal momenta of partial waves by values $\Delta\bm\pi_\bot=\bm\pi_\bot-\bm\pi_\bot^{(0)}$ which vary for different partial waves. These values seems to be rather small as compared with the kinetic energies of the waves ${\cal E}=\sqrt{m^2+\bm\pi^2}$. The corrections to the kinetic and total energies are defined by
\begin{equation}
{\cal E}={\cal E}_0+\frac{\bm\pi_\bot^{(0)}\cdot\Delta\bm
\pi_\bot}{{\cal E}_0^2}+\frac{\Delta\bm\pi_\bot^2}{2{\cal E}_0},\qquad {\cal E}_0=\sqrt{m^2+\bm\pi_\parallel^2+\left(\bm\pi_\bot^{(0)}\right)^2}.
\label{elftm}
\end{equation}

Following the discussion at the end of Sec. \ref{OAMinHIC}, we conclude that the both relations, $|\bm\pi_\bot^{(0)}|/{\cal E}_0$ and $|\Delta\bm
\pi_\bot|/{\cal E}_0$, are small. Therefore, the corrections to the energies of partial waves are rather small. We suppose that so small decoherence neither can lead to an important beam spread nor can prevent a detection of beam OAMs.

For photons which are mostly produced after the decay of the magnetic field, however, its presence is not important for our investigations. The search of OAMs of charged particles is effective when these particles are produced before of the above-mentioned decay.

It is well-known that the paraxial LG beams in a \emph{uniform} magnetic field are nonspreading. Nonparaxial twisted (and untwisted) beams can be considered as a result of some Lorentz boosts of LG beams \cite{Manipulating,photonPRA} while other approaches are also possible \cite{Karlovets2021,Karlovets2018}. Evidently, Lorentz boosts cannot lead to the beam spread.
Beams of scalar and vector mesons also possess above-mentioned properties.

A very important problem is a determination of appropriate methods for the detection of OAMs and a confirmation of a production of twisted particles in HIC. 
Twisted particles in experiments are always generated in a variety of OAMs.  
Application of OAM states crucially relies on the capability to detect these states 
accurately. The method of detecting or measuring OAM states is mostly based on the 
principle of interference or diffraction of waves, as the inherent properties of twisted 
particles. The experimental demonstration for detection of vortex synchrotron 
radiations has been well established, for instance, at the UVSOR-III electron storage 
ring \cite{experimental}.
In realistic applications, however, one may need to separate an individual OAM state 
even at the single particle level. In optics, this has been realized with many sophisticated 
designs, such as forked holograms \cite{Mair:2001}, Mach-Zehdner interferometer \cite{Leach:2004}, OAM sorter 
by log-polar optical transformation \cite{Berkhout:2010,Mirhosseini}, tunable resonator \cite{Wei2020}, etc. Similar devices 
have also been successfully applied in measuring or sorting OAM states for twisted 
electron beam using holograms or specific electrodes \cite{Grillo:2017,McMorran:2016lep,Suprano:2021qbe,Tavabi:2019lnd}. 
Although there are already well-developed devices measuring OAM states for 
conventional optical or electron vortex beams, the measurements to the twisted 
particles generated in HIC may still be of challenge. Conventional devices, for 
example, the optical holographic phase plates and OAM sorter cannot work with high 
energy photon in MeV region. Moreover, HIC is a rather non-trivial system, the 
generated ultra-strong magnetic field and rotation cannot be well-controlled. As a 
result, the detection of twisted particles created in HIC could be much more complicated than in
normal optical or electron vortex beams. As recently discussed in \cite{Maruyama:2022con}, the radiated 
photons in extremely strong magnetic fields are more likely to be non-coherent. This 
means former devices cannot be used in identification of such twisted photons. 
Fortunately, it has been shown that in Compton scattering of twisted photons with 
electrons the angular distribution and the polarization of the scattered photons depend 
non-trivially on the OAM and the opening angle of the incident vortex beam \cite{Maruyama:2017ptl}. Therefore, 
this quantum phenomenon can be used to 
measure the OAM of a twisted state applying detectors based on Compton scattering \cite{Maruyama:2022con}. 
As interests to twisted particles in particle collision physics increases, 
more and more quantum processes are proposed as alternatives for detecting or 
measuring the OAM states. At the same time, twisted states 
add a new degree of 
freedom to particle physics. It has been shown that 100 \% polarized vector mesons 
could be generated in a unpolarized twisted electron positron annihilation \cite{Ivanov:2022jzh,OAMinteraction}. This 
indicates that there could be a different possibility to explain the global polarization of 
various particles in HIC. Moreover, very recently authors in \cite{Fukushima:2020ncb} presented a very 
interesting result about mode decomposed chiral magnetic effect (CME) and chiral 
vortical effect (CVE) characterized in terms of quantum number $\ell$, the OAM appeared in the wave-function of fermion fields, i.e., the twisted states.
It is shown that the mode decomposed vector current is connected with the
chirality density by a simple relation. This can potentially have a close connection to 
the physics in HIC. Nevertheless, we would emphasize that
this discussion is still far from real measurements. We leave some details for future investigations.

In summary, we have investigated the possible production of twisted particles due to the inherent high vorticity and strong magnetic fields in HIC. We have rigorously proven that photons emitted at noncentral HIC due to electromagnetic interactions are highly twisted. We have also unambiguously shown that charged particles are produced in this case in nonspreading multiwave states and have significant OAMs. We expect that an emission of other particles in twisted states is rather ubiquitous. It should be noted that twisted particles manifest themselves in specific effects.

\begin{acknowledgments}
The authors are grateful to O.V. Teryaev for multiple discussions of the problem and valuable comments.
The work was supported by the National Natural Science
Foundation of China (grants No. 12175320, 11975320 and No. 11805242), the Natural Science Foundation of Guangdong Province, China (grant No. 2022A1515010280), by the Chinese Academy of Sciences President's International Fellowship Initiative (grant No. 2019VMA0019), and by the Fundamental Research Funds for the Central Universities, Sun Yat-sen University.
A. J. S. also acknowledges hospitality and support by the
Institute of Modern Physics of the Chinese Academy of Sciences.
\end{acknowledgments}


\begin{thebibliography}{99}
\bibitem{footnote}
The authors are indebted to O.V. Teryaev for clarification of this point and suggestion to consider twisted states in heavy-ion collisions as a counterpart of classical vorticity.
\bibitem{ffootnote}
In our paper, ``basic'' and ``unbasic'' states denote quantum states eligible and noneligible for a particle creation in a magnetic field. Only basic states are useful in statistical physics.
\end{thebibliography}
\end{document}